\documentclass[%
 preprintnumbers,
 amsmath,amssymb,
 prb, 
 twocolumn]{revtex4-2}

\usepackage{SIunits}
\usepackage{graphicx}
\usepackage{xspace}
\usepackage{braket}
\usepackage{bm}
\usepackage{placeins}
\usepackage{leftidx}
\usepackage{booktabs}
\usepackage{amsmath}
\usepackage{multirow}
\usepackage{longtable}
\usepackage[utf8]{inputenc}

\usepackage[usenames,dvipsnames]{color}

\usepackage[pdftex, colorlinks=true, urlcolor=blue, linkcolor=blue, citecolor=blue, pdfstartview=FitH, plainpages=false]{hyperref}

\newcommand{\kb}{\mathbf{k}}
\newcommand{\qb}{\mathbf{q}}
\newcommand{\qq}{\mathbf{q}}
\newcommand{\kk}{\mathbf{k}}

\newcommand{\enk}{\varepsilon_{n\kb}}
\newcommand{\emkq}{\varepsilon_{m\kb+\qb}}

\newcommand{\wnuq}{\omega_{\qb\nu}}
\newcommand{\qnu}{{{\qb\nu}}}
\newcommand{\wqnu}{{\omega_{\qnu}}}

\newcommand{\vnk}{\mathbf{v}_{n\kb}}
\newcommand{\vmkq}{\mathbf{v}_{m\kb+\qq}}
\newcommand{\vnka}{\mathrm{v}_{n\kb,\alpha}}
\newcommand{\vnkb}{\mathrm{v}_{n\kb,\beta}}

\newcommand{\abinit}{\textsc{Abinit}\xspace}

\newcommand{\perturbo}{\textsc{Perturbo}\xspace}
\newcommand{\amset}{\textsc{Amset}\xspace}
\newcommand{\boltztrap}{\textsc{Boltztrap}\xspace}

\newcommand{\gkkp}{g_{mn\nu}(\kb,\qb)}

\newcommand{\abinitio}{\textit{ab initio}\xspace}
\newcommand{\insilico}{\textit{in silico}\xspace}

\newcommand{\ie}{{\emph{i.e.}}\xspace}

\newcommand{\KS}{{\text{KS}}}

\newcommand{\BZ}{{\text{BZ}}}

\newcommand{\serta}{{\text{SERTA}}}
\newcommand{\crta}{{\text{CRTA}}}
\newcommand{\mrta}{{\text{MRTA}}}

\newcommand{\nk}{{n\kk}}
\newcommand{\mkq}{{m\kb+\qb}}

\newcommand*{\figrefletter}[2][]{\hyperref[{#2}]{#1}}

\begin{document}

\title{Assessing the quality of relaxation-time approximations with fully-automated computations of phonon-limited mobilities}
\date{\today}
\author{Romain Claes$^1$}
\author{Guillaume Brunin$^1$}
\author{Matteo Giantomassi$^1$}
\author{Gian-Marco Rignanese$^1$}
\author{Geoffroy Hautier$^{1,2}$}
\email[Corresponding author:~]{geoffroy.hautier@dartmouth.edu}

\affiliation{$^1$UCLouvain, Institute of Condensed Matter and Nanosciences (IMCN), Chemin des \'Etoiles~8, B-1348 Louvain-la-Neuve, Belgium}
\affiliation{$^2$Thayer School of Engineering, Dartmouth College, Hanover, New Hampshire 03755, USA}


\begin{abstract}


The mobility of carriers, as limited by their scattering with phonons, can now routinely be obtained from first-principles electron-phonon coupling calculations.
However, so far, most computations have relied on some form of simplification of the linearized Boltzmann transport equation based on either the self-energy, the momentum- or constant relaxation time approximations.
Here, we develop a high-throughput infrastructure and an automatic workflow and we compute 69 phonon-limited mobilities in semiconductors.
We compare the results resorting to the approximations with the exact iterative solution.
We conclude that the approximate values may deviate significantly from the exact ones and are thus not reliable.
Given the minimal computational overhead, our work encourages to rely on this exact iterative solution and warns on the possible inaccuracy of earlier results reported using relaxation time approximations.

\end{abstract}

\maketitle


Carrier mobility is one of the most important characteristics of semiconductor materials and
is a key property in many applications such as transistors, solar cells, thermoelectrics, transparent conductors, and light-emitting devices. 
Having reliable and efficient methods for computing these properties from first principles 
is therefore of crucial importance to improve the predictive power of \insilico design methods. 
Despite significant progress in the field since the late 2000s, 
very few bulk semiconductors have been investigated so far 
due to the complexity of the problem associated with the inherent large computational cost. 
Only recently, efficient implementations such as EPW~\cite{Giustino2007,Ponce2016}, 
\perturbo~\cite{zhou2021perturbo}, 
and \abinit~\cite{brunin2020phonon,brunin2020electron,Gonze2020,Romero2020}, 
have allowed the community to report \abinitio phonon-limited mobilities 
for a small set of semiconductors~\cite{ponce2020first}. 
These methods are all based on density-functional 
theory (DFT) for ground-state properties
and density-functional perturbation theory (DFPT) 
for vibrational properties~\cite{Gonze1997,Baroni2001}.
They all rely on Fourier-based interpolation schemes for the electron-phonon (e-ph) 
coupling matrix elements
as the microscopic description of e-ph scattering processes requires 
a very dense sampling of the whole Brillouin zone (BZ) that is unrealistic for DFPT calculations.
Recently, another methodology has been proposed with \amset~\cite{ganose2021efficient}
where low-cost first-principles inputs are used 
to define models for acoustic deformation potential, piezoelectric and polar optical phonon scattering. 
This approach effectively enables faster evaluations of phonon-limited mobilities
although not all the microscopic processes are explicitly accounted for with \abinitio quality.

In all these state-of-the-art methods, transport properties are usually computed within 
the linearized Boltzmann transport equation (BTE), 
where carriers are described in terms of wave packets that propagate
according to the semi-classical equations of motion between two consecutive scattering events.  
In principle, a fully \abinitio description of transport phenomena
should take into account several different scattering channels due to the interaction with, e.g., 
phonons, electrons, crystalline defects, ionized impurities, 
and even grain-boundaries in the case of polycrystalline samples.
In practice, most of the studies reported so far have focused on the
contribution due to e-ph scattering, while 
other effects such as the scattering by ionized impurities have been introduced 
either in an indirect way via semi-empirical models~\cite{ponce2018towards,ganose2021efficient} or via additional \abinitio calculations~\cite{Lu2019,Lu2020,lu2022first}.
At the level of the e-ph scattering, the predictive power of \abinitio methods 
has recently been quantified by estimating the impact of various effects such as 
spin-orbit coupling or many-body corrections to the e-ph vertex or 
effective masses in the case of silicon~\cite{ponce2018towards}. 
Even though a fully-microscopic theory of transport properties should 
take all these effects into account, 
it should also be noted that such corrections are usually applied within 
a given approximation to the BTE, usually based on a relaxation time ansatz. 
However, the appropriateness of the popular approximations to the BTE has  
not yet been quantified in real systems to the best of our knowledge. 
Here, we assess their quality 
in a quantitative and systematic way
using a computational workflow that allows us to investigate 69 phonon-limited mobilities among 
56 semiconductors under different relaxation-time approximations in a completely automated way.

The derivation of the BTE can be found in many publications such as Refs.~\cite{ponce2020first,ponce2018towards,Ziman1972,Blatt1957}.
In what follows, 
we summarize the most important equations 
and focus on electrons for the sake of brevity. 
Similar expressions can be easily obtained for holes.
Atomic units are used everywhere unless stated otherwise.
In semiconductors, the electron mobility ($\bm{\mu}_e$) 
is obtained by renormalizing the conductivity by the carrier 
concentration ($n_e$) 
and considering only the states in the conduction band (CB): 
\begin{equation}
    \mu_{e,\alpha\beta} = \frac{-1}{\Omega n_e}
    \sum_{n \in \text{CB}} \int_\BZ \frac{d\kb}{\Omega_\BZ}  \frac{\partial f_\nk}{\partial {\cal{E}}_\beta} \vnka,
    \label{eq:mobility_ibte}
\end{equation}
with $\alpha$ and $\beta$ the Cartesian coordinates,
$\Omega$ the volume of the unit cell, 
$\Omega_\BZ$ the volume of the first Brillouin zone (BZ), 
$n$ the electronic band index, $\kk$ the wave vector, 
$f_\nk$ the occupation of the state in the steady configuration,
$\bm{{\cal{E}}}$ the electric-field, 
and $\vnk$ the group velocity given by $\vnk = \nabla_{\kb}\varepsilon_{\nk}$,
with $\enk$ the band structure of the material. 
%
If we limit the discussion 
to drift mobilities where carriers are moving under the 
action of an electric field only 
(to the contrary of the Hall mobilities that also includes the effect of a magnetic field), 
the $\partial f_\nk/\partial {\cal{E}}_\beta$ term 
entering Eq.~\eqref{eq:mobility_ibte} 
is the solution of the following linear integral equation
%
\begin{equation}
    \begin{split}
        \frac{\partial f_\nk}{\partial {\cal{E}}_\beta} & = \frac{\partial f_\nk^0}{\partial \varepsilon} \text{v}_{\nk,\beta} \tau_\nk^0 
        + 2 \pi \tau_\nk^0 \sum_{m,\nu} \int_\BZ \frac{d\qb}{\Omega_\BZ} |\gkkp|^2 \\
                          & \times \left[ (n_\qnu^0 + f_\nk^0) \delta(\enk - \emkq  - \wnuq) \right.\\
                          & \left. + (n_\qnu^0 + 1 - f_\nk^0) \delta(\enk - \emkq  + \wnuq ) \right] \frac{\partial f_\mkq}{\partial {\cal{E}}_\beta},
    \end{split}
    \label{eq:bte}   
\end{equation}
with $\tau_\nk^0$, 
that has the dimension of time, 
given by
\begin{equation}
    \begin{split}
    \frac{1}{\tau^{0}_{n\kb}} = & 2 \pi \sum_{m,\nu} \int\frac{d\qb}{\Omega_\BZ}|\gkkp|^2\\
    & \times \left[ (n_\qnu^0 + f_{m\kb+\qb}^0)\delta(\enk - \emkq  + \wnuq)\right.\\
    & \left. + (n_\qnu^0 + 1 - f_{m\kb+\qb}^0)\delta(\enk - \emkq  - \wnuq ) \right].
    \end{split}
    \label{eq:lifetimes_bte}
\end{equation}
In Eqs.~\eqref{eq:bte}-\eqref{eq:lifetimes_bte}, 
$f_\nk^0$ is the equilibrium Fermi-Dirac occupation function,
$n_\qnu^0$ is the Bose-Einstein distribution for the phonon of wave vector $\qq$,
mode index $\nu$ and frequency $\wqnu$,
and $\gkkp$ are the e-ph coupling matrix elements
defined as 
$\gkkp = \braket{\psi_{m\kb+\qb}|\Delta_\qnu V^\KS|\psi_{n\kb}}$,
where $\psi_{n\kb}$ and $\psi_{m\kb+\qb}$
are the Kohn-Sham (KS) Bloch states and $\Delta_\qnu V^\KS$ is the phonon-induced variation of the self-consistent 
KS potential~\cite{brunin2020electron,brunin2020phonon,Giustino2017}.

Once discretized in the BZ, 
Eq.~\eqref{eq:bte} becomes a system of linear equations that is usually solved via iterative solvers (IBTE)
that take advantage of the 
sparsity of the scattering operator to reduce both memory and computational cost.
The main difficulty encountered when
solving Eq.~\eqref{eq:bte} is that very fine 
homogeneous $\kk$-point meshes are required 
to account for the coupling among all the $\partial f_{n\kk}/\partial {\cal{E}}_\beta$ terms and to properly converge transport properties.  
For this reason,
in many studies the full solution of the BTE 
is usually replaced by the so-called
self-energy relaxation time approximation (SERTA), 
where the second term on the right-hand side of Eq.~\eqref{eq:bte} 
is completely neglected. 
This leads to an explicit expression for 
$\partial f_\nk/\partial {\cal{E}}_\beta$ 
in terms of other quantities readily available.
In physical terms, this simplification corresponds to the 
relaxation time approximation (RTA)
in which carriers are assumed to
relax to the equilibrium Fermi-Dirac distribution $f_\nk^0$
with an exponential law and time constant $\tau^0_\nk$ 
when the external fields (electric and/or magnetic) are switched off.
Inserting this approximated solution in 
Eq.~\eqref{eq:mobility_ibte} gives:
\begin{equation}
    \mu_{e,\alpha\beta}^\serta = \frac{-1}{\Omega n_e}
    \sum_{n \in \text{CB}} \int_\BZ \frac{d\kb}{\Omega_\BZ} \frac{\partial f_\nk^0}{\partial \enk} \vnka \vnkb \tau_\nk^0.
    \label{eq:mobility_serta}
\end{equation}
The SERTA acronym stems from the fact that
$\tau_\nk^0$ is related to the inverse of the imaginary part of the e-ph 
Fan-Migdal self-energy~\cite{Giustino2017,brunin2020phonon} 
that gives the lifetime of a \emph{charged} quasi-particle excitation 
due to e-ph interactions.
In other words, the SERTA uses a relaxation-time ansatz 
to approximate the true solution of Eq.~\eqref{eq:bte} 
and assumes the transport relaxation time to be equal to 
the lifetime of a charged excitation. 
This is a reasonable but not necessarily correct assumption, 
especially because neglecting the second term 
on the right-hand side of Eq.~\eqref{eq:bte} 
corresponds to ignoring all the processes in which 
carriers are scattered back into the state $\nk$~\cite{ponce2020first}. 
%
%
%
In more geometrical words, the SERTA always underestimates the mobility 
because it does not differentiate between forward and backward scattering 
even though forward scattering (small angle between $\mathbf{v}_{\kb}$ and $\mathbf{v}_{\kb+\qb}$) 
does not deteriorate the mobility as much as backward scattering~\cite{ponce2021first}.
In order to partially account for these back-scattering events, 
the momentum-relaxation time approximation (MRTA) has been introduced 
where the integrand in Eq.~\eqref{eq:lifetimes_bte} is now weighted by 
the efficiency factor~\cite{Ma2018,Li2015,ganose2021efficient,ponce2020first}
\begin{equation}
    \alpha_{mn}^\mrta(\kk,\qq) = 
    \left(1-\frac{\vnk.\vmkq}{|\vnk|^2} \right).
    \label{eq:efficiency_factor}
\end{equation}
Indeed, this favors forward scattering geometrically and therefore allows to 
better take into account the relative changes 
in the electron velocity due to the scattering processes. 
Finally, an even more crude approximation consists in 
assuming the relaxation times to be constant for all $\nk$, 
\ie $\tau_\nk = \tau$ in Eq.~\eqref{eq:mobility_serta}. 
In this constant relaxation time approximation (CRTA), 
the mobility depends only on the band structure 
and on $\tau$ as a phenomenological parameter.
Since the computation of phonons and
e-ph matrix elements are not needed, 
this greatly reduces the computational cost.
This is the main reason why
the CRTA has been widely employed in the past.
In particular, it has been implemented in \boltztrap~\cite{Madsen2006,Madsen2018}
and extensively used, for instance to rank large sets of materials 
using a common lifetime~\cite{hautier2013identification,ricci2017ab,ricci2020gapped,faghaninia2017computational} 
(or, equivalently, by looking at the transport effective mass~\cite{brunin2019transparent,gibbs2017effective}). 
Another application lies in understanding experimental results 
by determining the lifetime in a specific material~\cite{smiadak2022quasi,chmielowski2015theoretical,Madsen2018}.


All these expressions have been implemented 
in the transport module of \abinit. 
Our approach, detailed in Ref.~\cite{brunin2020phonon},
takes advantage of (i) the tetrahedron integration scheme to 
reduce the number of e-ph transitions to be computed, (ii)
a Fourier interpolation of the scattering potentials in $\qq$ space 
including the proper treatment of dipole and quadrupole contributions, and
(iii) exact KS wavefunctions that are computed only for the $\kk$-points lying inside a small energy window around the band edges.
This procedure allows us to bypass the generation of maximally-localized Wannier functions, 
a major advantage in our context as all the steps of our workflow can be easily automated.

Even though the CRTA, SERTA, and MRTA have been widely used in the literature 
to characterize the transport of electrons or holes, the validity of such 
approximations has not yet been tested in a systematic way. 
Here, we analyze the phonon-limited transport in 56 different semiconductors, 
including 56 electron and 13 hole mobilities. 
This allows us to directly probe the quality of these different approximations to
the IBTE. 
These 56 semiconductors have been selected by the following procedure. 
In order to reduce the total computation time, 
we consider only those semiconductors for which phonon properties are available in the Materials Project database~\cite{Petretto2018,jain2013commentary} 
and discard materials with imaginary phonon frequencies (vibrational instabilities) or those that are predicted to be thermodynamically unstable, 
\ie with an energy above hull larger than 50~meV/atom. 
The results of Ref.~\cite{ricci2017ab} have then been used 
to remove all materials for which 
the average transport effective mass is larger than one.
%
Finally, we enforce two additional constraints 
that are needed in order to be compatible 
with the previous DFPT calculations performed with PBEsol scalar-relativistic norm-conserving pseudopotentials including
non-linear core correction (NLCC)~\cite{van2018pseudodojo}.
First, 
we have considered systems with a single conduction/valence band
within an energy window of 0.25~eV 
above/below the minimum/maximum.
The motivation is that spin-orbit coupling (SOC) has been shown to have a significant impact on phonon-limited mobilities~\cite{ponce2021first}. 
%
Restricting our database to systems with a single band 
allows us to avoid the worst-case scenario of degenerate hole states that are split by SOC
although it is clear that a proper treatment of relativistic effects in mobility calculations would require the inclusion of SOC effects both at the electronic and vibrational level.
%
Secondly we have considered only space groups 
for which the dynamical quadrupoles $Q^*$
are zero by symmetry~\footnote{Dynamical quadrupoles are non-zero in 
all non-centrosymmetric crystals, 
but also in centrosymmetric ones 
if one or more atoms are placed at non-centrosymmetric sites.}.
As recently shown in Refs.~\cite{brunin2020electron,brunin2020phonon,jhalani2020piezoelectric,park2020long,ponce2021first}, 
dynamical quadrupoles play a crucial role 
for obtaining reliable phonon-limited mobilities in semiconductors.
Unfortunately, the DFPT computation of
$Q^*$ 
is presently limited to norm-conserving pseudopotentials 
without NLCC, hence we decided to restrict the discussion to high-symmetry structures.
Overall, our screening criteria led to $56$ materials 
($37$ in the Fm$\bar{3}$m space group, $18$ in Pm$\bar{3}$m
and one belonging to the tetragonal P4/mmm space group)
and $69$ mobilities ($56$ electron and $13$ hole mobilities). 
Although our dataset mostly consists of cubic systems,
we expect our analysis to hold for other structures as well.

\begin{figure}
    \centering
    \includegraphics[scale=0.245]{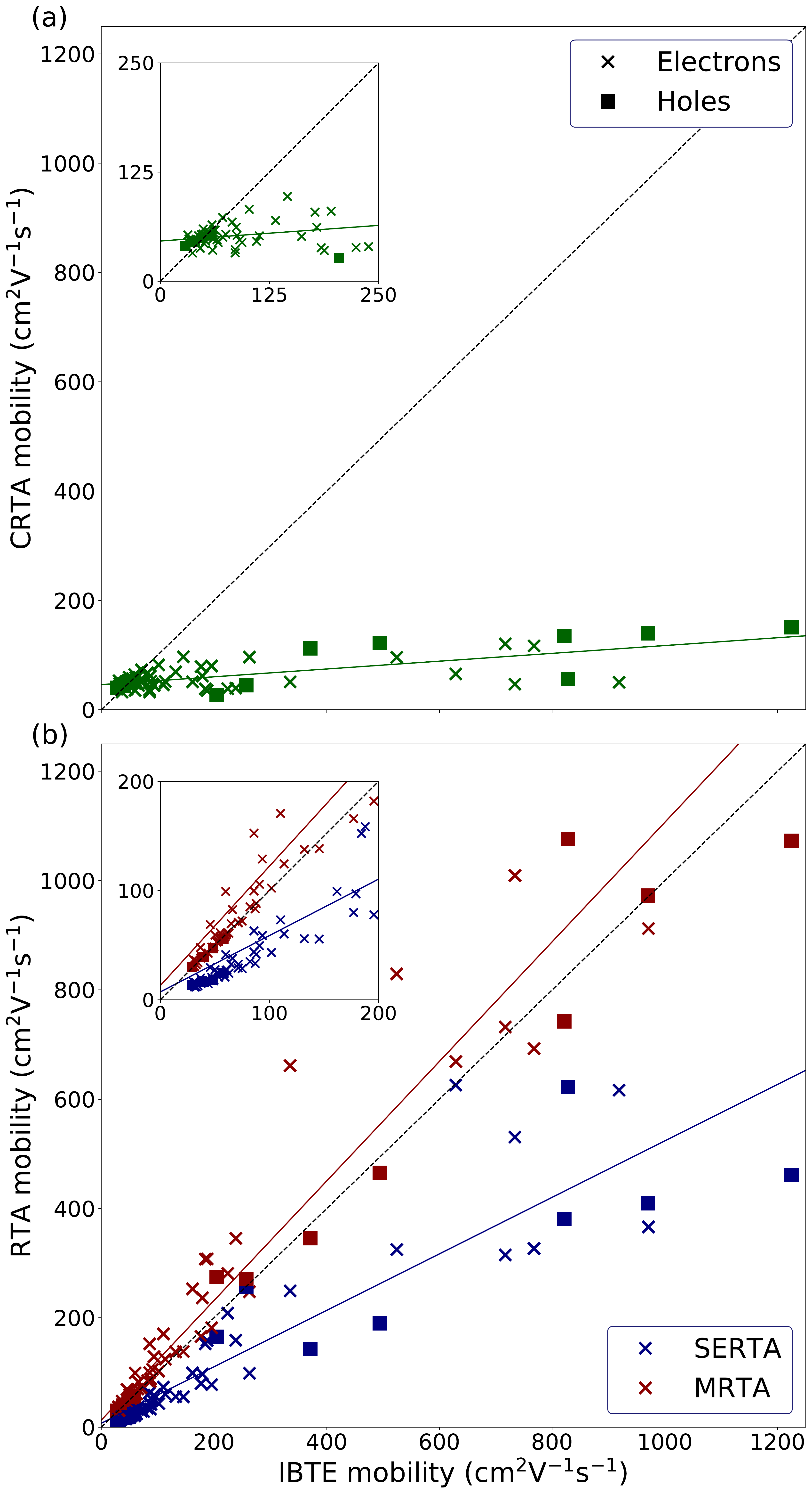}
    \caption{Comparison of the (a) CRTA, and (b) SERTA, MRTA with the IBTE electron mobilities. 
    For the CRTA, the chosen lifetime minimizes the mean absolute error. 
    MRTA mobilities are in red whereas SERTA mobilities are in blue. 
    The black dotted lines represent the IBTE results.
    The green, blue and red solid lines are linear fits 
    of the CRTA, SERTA and MRTA results, respectively.}
    \label{fig:approx_methods}
\end{figure}

Using our automated mobility computations, 
we can compare the SERTA, MRTA and CRTA results 
versus the exact IBTE ones in a systematic way. 
Figure~\ref{fig:approx_methods} 
shows this comparison of the CRTA \figrefletter[(a)]{fig:approx_methods} and 
SERTA/MRTA \figrefletter[(b)]{fig:approx_methods}  mobilities 
with the IBTE results for all the systems in our dataset. 
The numerical results can be found in Table S1 of the Supplemental Material~\cite{supplemental}.
Obviously, the CRTA mobility for a given material
can be made exactly equal to the IBTE value by an appropriate choice of  $\tau$.
In our dataset, a wide variety of lifetimes would have to be used ranging
from $6$~fs for CsCaBr$_3$ to $188$~fs for KMgH$_3$. 
Considering the complete set of materials, a lifetime of $10.3$~fs minimizes the mean absolute percentage error but, 
as can be seen from Fig.~\ref{fig:approx_methods}\figrefletter[(a)]{fig:approx_methods}, 
the agreement with the IBTE is only valid for low IBTE mobilities, 
as most of the systems in the dataset have an IBTE mobility lower than 100~cm$^2$V$^{-1}$s$^{-1}$ (see the insets of Fig.~\ref{fig:approx_methods}). 
However, even in this region, a material with a low CRTA mobility may show a large relative error 
(see Fig.~S1 of the Supplemental Material~\cite{supplemental}).
It is clear from Fig.~\ref{fig:approx_methods}\figrefletter[(a)]{fig:approx_methods} 
that the correlation is weak between the CRTA and IBTE results. 
A large (low) CRTA result does not guarantee a large (low) IBTE mobility.
One can also compute Spearman's rank correlation coefficient $\rho$ 
in order to quantify the ranking capability of CRTA. 
A value of $\rho^\crta = 0.44$ is obtained, 
indicating that the CRTA is overall not able 
to correctly rank materials from our dataset. 
%
%
This analysis therefore highlights the importance of going beyond the CRTA for accurate results. 
In particular, in material screening or high-throughput computing where this approach has been very popular, 
this indicates that the CRTA mobility or, equivalently, 
the transport effective mass~\cite{ricci2017ab,brunin2019transparent}, 
should 
be used with extreme caution as a first filter to identify materials with high mobility
and should be followed by a higher-level analysis of the transport if possible. 

It is clear from Fig.~\ref{fig:approx_methods}\figrefletter[(b)]{fig:approx_methods} that the SERTA and MRTA both perform better than the CRTA. 
In particular, in terms of ranking materials of our dataset, 
both of them prove to be adequate, with $\rho^\serta = 0.97$ and $\rho^\mrta = 0.98$. 
Additionally, Fig.~\ref{fig:approx_methods}\figrefletter[(b)]{fig:approx_methods} shows that 
the MRTA performs overall better than the SERTA 
in approximating the IBTE, 
with a mean absolute percentage error of 18\% 
for the former and 48\% for the latter. 
In the materials investigated here, the MRTA mobility is always higher than the 
one predicted by the SERTA.
Indeed, we specifically selected materials with a single band, that is often 
located around $\Gamma$ in the BZ. In this case, intravalley scattering largely dominates, 
in particular with small wave vectors $\qq$ because the effective masses are lower than 1, 
hence the bands are relatively dispersive~\cite{brunin2020phonon}. 
Note that intravalley scattering plays a crucial role 
even in cases where the band extrema are not at $\Gamma$. 
If the important wave vectors $\qq$ are small, then $\alpha^\mrta$ is between 0 and 1, 
and the mobility increases from the SERTA to the MRTA. 
However, when compared to the IBTE, 
there is no general rule for the MRTA and it can underestimate or 
overestimate the mobility. 

In the literature, SERTA has recently emerged as the most satisfactory approach. 
Indeed, for the few systems investigated so far (such as 
Si~\cite{brunin2020phonon, ponce2018towards, Ma2018} 
or GaAs~\cite{brunin2020phonon, zhou2016ab}), it is predicted that
the SERTA mobilities are closer to experimental data than the MRTA or IBTE results.
Indeed, in Si, the differences between the SERTA, MRTA, and IBTE mobilities are lower than 5\%, 
all of them being very close to experimental data~\cite{ponce2018towards,brunin2020phonon,Ma2018}. 
However, in GaAs, there is a large spread in the reported computed mobilities, 
which can partly be explained by the different transport formalisms 
used for the computations, 
since the SERTA (MRTA) underestimates (overestimates) 
the IBTE solution by 52\% (4\%). 
From Fig.~\ref{fig:approx_methods} and Table S1, it is clear that
some outliers show important deviations from the IBTE mobility. 
For instance, in SrO, KH, KMgH$_3$ and MgO, 
the MRTA leads to errors larger than 60\% 
when compared to the exact IBTE solution. 
From a computational point of view, 
it is difficult to estimate whether the MRTA will 
lead to a result comparable to the IBTE or not 
without actually doing both computations. 
This indicates that both the SERTA and MRTA are actually unreliable and 
the IBTE should always be preferred. 
In particular, before comparing computed mobilities to experimental data, 
we believe it is crucial to first make sure 
that any approximation used in the process is reliable,
or at least to quantify the error on the final quantity. 
We point out that using spatial and time-reversal symmetries, 
and a similar filtering method as presented in Ref.~\cite{brunin2020phonon},
solving the IBTE can be seen as a post-processing of the SERTA 
that does not lead to a significant 
increase of computational time nor memory. 
It is also worth noting that 
the type of charge carrier does not seem to affect these results. 

The computed IBTE mobilities overestimate the experimental data, 
which is expected since other sources of scattering (e.g., impurity scattering) are completely ignored. 
In addition, most of the experimental mobilities reported
in the literature are measured using the Hall effect and 
it is therefore necessary to weight computational results by a 
material-dependent Hall factor that typically ranges 
between $0.7$ and $2$~\cite{ponce2021first}.
The inclusion of the Hall mobility in \abinit is left for future work. 


The computation of the phonon-limited mobility is a rather 
complex task involving many steps, 
that typically require an important human time and intervention. 
To fully automate all the different parts of the computation, 
including the convergence studies for the BZ sampling,
we have developed a workflow within the AbiPy python package~\cite{abipy-website}. 
The main steps are schematically represented in Fig.~\ref{fig:workflow}, 
with more details given in the Supplemental Material~\cite{supplemental}.
\begin{figure}
    \centering
    \includegraphics[width=0.48\textwidth]{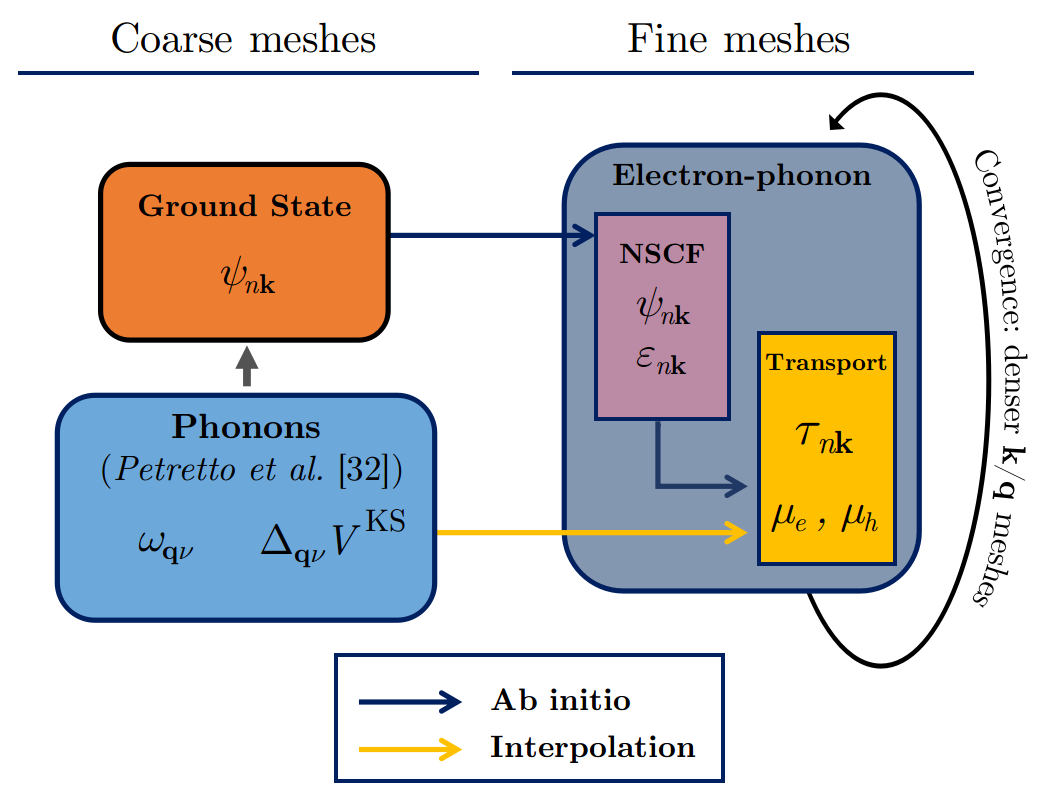}
    \caption{Flowchart illustrating the workflow used to automatically compute phonon-limited mobilities.}
    \label{fig:workflow}
\end{figure}
The ingredients needed in Eq.~\eqref{eq:bte} are 
the KS wave functions on the dense mesh for the electronic part 
(in purple in Fig.~\ref{fig:workflow}) and 
the DFPT scattering potentials and the interatomic force constants 
on a coarse mesh (typical of DFPT) for the phonon part 
(in blue in Fig.~\ref{fig:workflow}). 
The latter can be easily computed with another AbiPy workflow,  
although in this study we prefer to start from 
a database of previous DFPT computations~\cite{Petretto2018,Petretto2018b}. 
The first step of the workflow consists in a ground-state calculation 
(in orange in Fig.~\ref{fig:workflow}), 
with basic parameters reused from the DFPT database.
This allows to determine the wave functions 
on the fine mesh using a two-step procedure described 
in Ref.~\cite{brunin2020phonon} and 
in the Supplemental Material~\cite{supplemental}. 
All the ingredients required to compute the mobility on a given dense mesh 
are then readily available.
Since a convergence study is needed with respect to this dense mesh,  
we perform the previous steps multiple times 
for meshes of increasing density. 
Convergence is assumed to be reached when three consecutive grids lead to 
mobilities maximum 5\% away from each other.



In conclusion, we have obtained well-converged phonon-limited mobilities 
for a rather large set of semiconductors.
We have developed and used an automatic workflow 
that allows for the comparison of different approximations to the BTE
with the exact results.
Our results show that both SERTA and MRTA   
are not reliable in general and that they are in many cases not good 
approximations to the IBTE. 
This should be kept in mind when looking at previously-published 
results using any of these approaches.
The fact that the SERTA/MRTA agree with experiments for a few systems 
is not sufficient to establish these methods as a standard. 
Even though the IBTE requires more computational power and implementation, 
it should be the recommended approach for computing transport properties.
Finally, our work demonstrates that phonon-limited mobilities can be computed automatically and high-throughput opening new avenues for materials screening.
%


\section{Acknowledgment}
G.-M.~R.~acknowledges financial support from F.R.S.-FNRS.
Computational resources have been provided by the Consortium des Équipements de Calcul Intensif (CÉCI), funded by the Fonds de la Recherche Scientifique de Belgique (F.R.S.-FNRS) under Grant No.~2.5020.11 and by the Walloon Region.
The present research benefited from computational resources made available on the {Tier-1} supercomputer of the F\'ed\'eration Wallonie-Bruxelles, infrastructure funded by the Walloon Region under grant agreement n\textsuperscript{o}1117545. G.H. acknowledges funding by the U.S. Department of Energy, Office of Science, Office of Basic Energy Sciences, Materials Sciences and Engineering Division, under Contract DE-AC02-05-CH11231: Materials Project program KC23MP.
%
%
\nocite{Shankland1971,Euwema1969,Koelling1986} 
\bibliographystyle{apsrev4-1}
\bibliography{Bibliography}

\begin{thebibliography}{44}%
\makeatletter
\providecommand \@ifxundefined [1]{%
 \@ifx{#1\undefined}
}%
\providecommand \@ifnum [1]{%
 \ifnum #1\expandafter \@firstoftwo
 \else \expandafter \@secondoftwo
 \fi
}%
\providecommand \@ifx [1]{%
 \ifx #1\expandafter \@firstoftwo
 \else \expandafter \@secondoftwo
 \fi
}%
\providecommand \natexlab [1]{#1}%
\providecommand \enquote  [1]{``#1''}%
\providecommand \bibnamefont  [1]{#1}%
\providecommand \bibfnamefont [1]{#1}%
\providecommand \citenamefont [1]{#1}%
\providecommand \href@noop [0]{\@secondoftwo}%
\providecommand \href [0]{\begingroup \@sanitize@url \@href}%
\providecommand \@href[1]{\@@startlink{#1}\@@href}%
\providecommand \@@href[1]{\endgroup#1\@@endlink}%
\providecommand \@sanitize@url [0]{\catcode `\\12\catcode `\$12\catcode
  `\&12\catcode `\#12\catcode `\^12\catcode `\_12\catcode `\%12\relax}%
\providecommand \@@startlink[1]{}%
\providecommand \@@endlink[0]{}%
\providecommand \url  [0]{\begingroup\@sanitize@url \@url }%
\providecommand \@url [1]{\endgroup\@href {#1}{\urlprefix }}%
\providecommand \urlprefix  [0]{URL }%
\providecommand \Eprint [0]{\href }%
\providecommand \doibase [0]{http://dx.doi.org/}%
\providecommand \selectlanguage [0]{\@gobble}%
\providecommand \bibinfo  [0]{\@secondoftwo}%
\providecommand \bibfield  [0]{\@secondoftwo}%
\providecommand \translation [1]{[#1]}%
\providecommand \BibitemOpen [0]{}%
\providecommand \bibitemStop [0]{}%
\providecommand \bibitemNoStop [0]{.\EOS\space}%
\providecommand \EOS [0]{\spacefactor3000\relax}%
\providecommand \BibitemShut  [1]{\csname bibitem#1\endcsname}%
\let\auto@bib@innerbib\@empty
\bibitem [{\citenamefont {Giustino}\ \emph {et~al.}(2007)\citenamefont
  {Giustino}, \citenamefont {Cohen},\ and\ \citenamefont
  {Louie}}]{Giustino2007}%
  \BibitemOpen
  \bibfield  {author} {\bibinfo {author} {\bibfnamefont {F.}~\bibnamefont
  {Giustino}}, \bibinfo {author} {\bibfnamefont {M.~L.}\ \bibnamefont {Cohen}},
  \ and\ \bibinfo {author} {\bibfnamefont {S.~G.}\ \bibnamefont {Louie}},\
  }\href {\doibase 10.1103/PhysRevB.76.165108} {\bibfield  {journal} {\bibinfo
  {journal} {Phys.~Rev.~B}\ }\textbf {\bibinfo {volume} {76}},\ \bibinfo
  {pages} {165108} (\bibinfo {year} {2007})}\BibitemShut {NoStop}%
\bibitem [{\citenamefont {Ponc{\'e}}\ \emph {et~al.}(2016)\citenamefont
  {Ponc{\'e}}, \citenamefont {Margine}, \citenamefont {Verdi},\ and\
  \citenamefont {Giustino}}]{Ponce2016}%
  \BibitemOpen
  \bibfield  {author} {\bibinfo {author} {\bibfnamefont {S.}~\bibnamefont
  {Ponc{\'e}}}, \bibinfo {author} {\bibfnamefont {E.~R.}\ \bibnamefont
  {Margine}}, \bibinfo {author} {\bibfnamefont {C.}~\bibnamefont {Verdi}}, \
  and\ \bibinfo {author} {\bibfnamefont {F.}~\bibnamefont {Giustino}},\ }\href
  {\doibase 10.1016/j.cpc.2016.07.028} {\bibfield  {journal} {\bibinfo
  {journal} {Comput.~Phys.~Commun.}\ }\textbf {\bibinfo {volume} {209}},\
  \bibinfo {pages} {116} (\bibinfo {year} {2016})}\BibitemShut {NoStop}%
\bibitem [{\citenamefont {Zhou}\ \emph {et~al.}(2021)\citenamefont {Zhou},
  \citenamefont {Park}, \citenamefont {Lu}, \citenamefont {Maliyov},
  \citenamefont {Tong},\ and\ \citenamefont {Bernardi}}]{zhou2021perturbo}%
  \BibitemOpen
  \bibfield  {author} {\bibinfo {author} {\bibfnamefont {J.-J.}\ \bibnamefont
  {Zhou}}, \bibinfo {author} {\bibfnamefont {J.}~\bibnamefont {Park}}, \bibinfo
  {author} {\bibfnamefont {I.-T.}\ \bibnamefont {Lu}}, \bibinfo {author}
  {\bibfnamefont {I.}~\bibnamefont {Maliyov}}, \bibinfo {author} {\bibfnamefont
  {X.}~\bibnamefont {Tong}}, \ and\ \bibinfo {author} {\bibfnamefont
  {M.}~\bibnamefont {Bernardi}},\ }\href {\doibase 10.1016/j.cpc.2021.107970}
  {\bibfield  {journal} {\bibinfo  {journal} {Comput.~Phys.~Commun.}\ }\textbf
  {\bibinfo {volume} {264}},\ \bibinfo {pages} {107970} (\bibinfo {year}
  {2021})}\BibitemShut {NoStop}%
\bibitem [{\citenamefont {Brunin}\ \emph
  {et~al.}(2020{\natexlab{a}})\citenamefont {Brunin}, \citenamefont {Miranda},
  \citenamefont {Giantomassi}, \citenamefont {Royo}, \citenamefont {Stengel},
  \citenamefont {Verstraete}, \citenamefont {Gonze}, \citenamefont
  {Rignanese},\ and\ \citenamefont {Hautier}}]{brunin2020phonon}%
  \BibitemOpen
  \bibfield  {author} {\bibinfo {author} {\bibfnamefont {G.}~\bibnamefont
  {Brunin}}, \bibinfo {author} {\bibfnamefont {H.~P.~C.}\ \bibnamefont
  {Miranda}}, \bibinfo {author} {\bibfnamefont {M.}~\bibnamefont
  {Giantomassi}}, \bibinfo {author} {\bibfnamefont {M.}~\bibnamefont {Royo}},
  \bibinfo {author} {\bibfnamefont {M.}~\bibnamefont {Stengel}}, \bibinfo
  {author} {\bibfnamefont {M.~J.}\ \bibnamefont {Verstraete}}, \bibinfo
  {author} {\bibfnamefont {X.}~\bibnamefont {Gonze}}, \bibinfo {author}
  {\bibfnamefont {G.-M.}\ \bibnamefont {Rignanese}}, \ and\ \bibinfo {author}
  {\bibfnamefont {G.}~\bibnamefont {Hautier}},\ }\href {\doibase
  10.1103/PhysRevB.102.094308} {\bibfield  {journal} {\bibinfo  {journal}
  {Phys.~Rev.~B}\ }\textbf {\bibinfo {volume} {102}},\ \bibinfo {pages}
  {094308} (\bibinfo {year} {2020}{\natexlab{a}})}\BibitemShut {NoStop}%
\bibitem [{\citenamefont {Brunin}\ \emph
  {et~al.}(2020{\natexlab{b}})\citenamefont {Brunin}, \citenamefont {Miranda},
  \citenamefont {Giantomassi}, \citenamefont {Royo}, \citenamefont {Stengel},
  \citenamefont {Verstraete}, \citenamefont {Gonze}, \citenamefont
  {Rignanese},\ and\ \citenamefont {Hautier}}]{brunin2020electron}%
  \BibitemOpen
  \bibfield  {author} {\bibinfo {author} {\bibfnamefont {G.}~\bibnamefont
  {Brunin}}, \bibinfo {author} {\bibfnamefont {H.~P.~C.}\ \bibnamefont
  {Miranda}}, \bibinfo {author} {\bibfnamefont {M.}~\bibnamefont
  {Giantomassi}}, \bibinfo {author} {\bibfnamefont {M.}~\bibnamefont {Royo}},
  \bibinfo {author} {\bibfnamefont {M.}~\bibnamefont {Stengel}}, \bibinfo
  {author} {\bibfnamefont {M.~J.}\ \bibnamefont {Verstraete}}, \bibinfo
  {author} {\bibfnamefont {X.}~\bibnamefont {Gonze}}, \bibinfo {author}
  {\bibfnamefont {G.-M.}\ \bibnamefont {Rignanese}}, \ and\ \bibinfo {author}
  {\bibfnamefont {G.}~\bibnamefont {Hautier}},\ }\href {\doibase
  10.1103/PhysRevLett.125.136601} {\bibfield  {journal} {\bibinfo  {journal}
  {Phys.~Rev.~Lett.}\ }\textbf {\bibinfo {volume} {125}},\ \bibinfo {pages}
  {136601} (\bibinfo {year} {2020}{\natexlab{b}})}\BibitemShut {NoStop}%
\bibitem [{\citenamefont {Gonze}\ \emph {et~al.}(2020)\citenamefont {Gonze},
  \citenamefont {Amadon}, \citenamefont {Antonius}, \citenamefont {Arnardi},
  \citenamefont {Baguet}, \citenamefont {Beuken}, \citenamefont {Bieder},
  \citenamefont {Bottin}, \citenamefont {Bouchet}, \citenamefont {Bousquet}
  \emph {et~al.}}]{Gonze2020}%
  \BibitemOpen
  \bibfield  {author} {\bibinfo {author} {\bibfnamefont {X.}~\bibnamefont
  {Gonze}}, \bibinfo {author} {\bibfnamefont {B.}~\bibnamefont {Amadon}},
  \bibinfo {author} {\bibfnamefont {G.}~\bibnamefont {Antonius}}, \bibinfo
  {author} {\bibfnamefont {F.}~\bibnamefont {Arnardi}}, \bibinfo {author}
  {\bibfnamefont {L.}~\bibnamefont {Baguet}}, \bibinfo {author} {\bibfnamefont
  {J.-M.}\ \bibnamefont {Beuken}}, \bibinfo {author} {\bibfnamefont
  {J.}~\bibnamefont {Bieder}}, \bibinfo {author} {\bibfnamefont
  {F.}~\bibnamefont {Bottin}}, \bibinfo {author} {\bibfnamefont
  {J.}~\bibnamefont {Bouchet}}, \bibinfo {author} {\bibfnamefont
  {E.}~\bibnamefont {Bousquet}},  \emph {et~al.},\ }\href {\doibase
  https://doi.org/10.1016/j.cpc.2019.107042} {\bibfield  {journal} {\bibinfo
  {journal} {Comput.~Phys.~Commun.}\ }\textbf {\bibinfo {volume} {248}},\
  \bibinfo {pages} {107042} (\bibinfo {year} {2020})}\BibitemShut {NoStop}%
\bibitem [{\citenamefont {Romero}\ \emph {et~al.}(2020)\citenamefont {Romero},
  \citenamefont {Allan}, \citenamefont {Amadon}, \citenamefont {Antonius},
  \citenamefont {Applencourt}, \citenamefont {Baguet}, \citenamefont {Bieder},
  \citenamefont {Bottin}, \citenamefont {Bouchet}, \citenamefont {Bousquet}
  \emph {et~al.}}]{Romero2020}%
  \BibitemOpen
  \bibfield  {author} {\bibinfo {author} {\bibfnamefont {A.~H.}\ \bibnamefont
  {Romero}}, \bibinfo {author} {\bibfnamefont {D.~C.}\ \bibnamefont {Allan}},
  \bibinfo {author} {\bibfnamefont {B.}~\bibnamefont {Amadon}}, \bibinfo
  {author} {\bibfnamefont {G.}~\bibnamefont {Antonius}}, \bibinfo {author}
  {\bibfnamefont {T.}~\bibnamefont {Applencourt}}, \bibinfo {author}
  {\bibfnamefont {L.}~\bibnamefont {Baguet}}, \bibinfo {author} {\bibfnamefont
  {J.}~\bibnamefont {Bieder}}, \bibinfo {author} {\bibfnamefont
  {F.}~\bibnamefont {Bottin}}, \bibinfo {author} {\bibfnamefont
  {J.}~\bibnamefont {Bouchet}}, \bibinfo {author} {\bibfnamefont
  {E.}~\bibnamefont {Bousquet}},  \emph {et~al.},\ }\href {\doibase
  https://doi.org/10.1063/1.5144261} {\bibfield  {journal} {\bibinfo  {journal}
  {J.~Chem.~Phys.}\ }\textbf {\bibinfo {volume} {152}},\ \bibinfo {pages}
  {124102} (\bibinfo {year} {2020})}\BibitemShut {NoStop}%
\bibitem [{\citenamefont {Ponc{\'e}}\ \emph {et~al.}(2020)\citenamefont
  {Ponc{\'e}}, \citenamefont {Li}, \citenamefont {Reichardt},\ and\
  \citenamefont {Giustino}}]{ponce2020first}%
  \BibitemOpen
  \bibfield  {author} {\bibinfo {author} {\bibfnamefont {S.}~\bibnamefont
  {Ponc{\'e}}}, \bibinfo {author} {\bibfnamefont {W.}~\bibnamefont {Li}},
  \bibinfo {author} {\bibfnamefont {S.}~\bibnamefont {Reichardt}}, \ and\
  \bibinfo {author} {\bibfnamefont {F.}~\bibnamefont {Giustino}},\ }\href
  {\doibase 10.1088/1361-6633/ab6a43} {\bibfield  {journal} {\bibinfo
  {journal} {Rep.~Prog.~Phys.}\ }\textbf {\bibinfo {volume} {83}},\ \bibinfo
  {pages} {036501} (\bibinfo {year} {2020})}\BibitemShut {NoStop}%
\bibitem [{\citenamefont {Gonze}\ and\ \citenamefont {Lee}(1997)}]{Gonze1997}%
  \BibitemOpen
  \bibfield  {author} {\bibinfo {author} {\bibfnamefont {X.}~\bibnamefont
  {Gonze}}\ and\ \bibinfo {author} {\bibfnamefont {C.}~\bibnamefont {Lee}},\
  }\href {\doibase 10.1103/PhysRevB.55.10355} {\bibfield  {journal} {\bibinfo
  {journal} {Phys.~Rev.~B}\ }\textbf {\bibinfo {volume} {55}},\ \bibinfo
  {pages} {10355} (\bibinfo {year} {1997})}\BibitemShut {NoStop}%
\bibitem [{\citenamefont {Baroni}\ \emph {et~al.}(2001)\citenamefont {Baroni},
  \citenamefont {de~Gironcoli}, \citenamefont {Dal~Corso},\ and\ \citenamefont
  {Giannozzi}}]{Baroni2001}%
  \BibitemOpen
  \bibfield  {author} {\bibinfo {author} {\bibfnamefont {S.}~\bibnamefont
  {Baroni}}, \bibinfo {author} {\bibfnamefont {S.}~\bibnamefont
  {de~Gironcoli}}, \bibinfo {author} {\bibfnamefont {A.}~\bibnamefont
  {Dal~Corso}}, \ and\ \bibinfo {author} {\bibfnamefont {P.}~\bibnamefont
  {Giannozzi}},\ }\href {\doibase 10.1103/revmodphys.73.515} {\bibfield
  {journal} {\bibinfo  {journal} {Rev.~Mod.~Phys.}\ }\textbf {\bibinfo {volume}
  {73}},\ \bibinfo {pages} {515} (\bibinfo {year} {2001})}\BibitemShut
  {NoStop}%
\bibitem [{\citenamefont {Ganose}\ \emph {et~al.}(2021)\citenamefont {Ganose},
  \citenamefont {Park}, \citenamefont {Faghaninia}, \citenamefont
  {Woods-Robinson}, \citenamefont {Persson},\ and\ \citenamefont
  {Jain}}]{ganose2021efficient}%
  \BibitemOpen
  \bibfield  {author} {\bibinfo {author} {\bibfnamefont {A.~M.}\ \bibnamefont
  {Ganose}}, \bibinfo {author} {\bibfnamefont {J.}~\bibnamefont {Park}},
  \bibinfo {author} {\bibfnamefont {A.}~\bibnamefont {Faghaninia}}, \bibinfo
  {author} {\bibfnamefont {R.}~\bibnamefont {Woods-Robinson}}, \bibinfo
  {author} {\bibfnamefont {K.~A.}\ \bibnamefont {Persson}}, \ and\ \bibinfo
  {author} {\bibfnamefont {A.}~\bibnamefont {Jain}},\ }\href {\doibase
  https://doi.org/10.1038/s41467-021-22440-5} {\bibfield  {journal} {\bibinfo
  {journal} {Nat.~Commun.}\ }\textbf {\bibinfo {volume} {12}},\ \bibinfo
  {pages} {1} (\bibinfo {year} {2021})}\BibitemShut {NoStop}%
\bibitem [{\citenamefont {Ponc{\'e}}\ \emph {et~al.}(2018)\citenamefont
  {Ponc{\'e}}, \citenamefont {Margine},\ and\ \citenamefont
  {Giustino}}]{ponce2018towards}%
  \BibitemOpen
  \bibfield  {author} {\bibinfo {author} {\bibfnamefont {S.}~\bibnamefont
  {Ponc{\'e}}}, \bibinfo {author} {\bibfnamefont {E.~R.}\ \bibnamefont
  {Margine}}, \ and\ \bibinfo {author} {\bibfnamefont {F.}~\bibnamefont
  {Giustino}},\ }\href {\doibase https://doi.org/10.1103/PhysRevB.97.121201}
  {\bibfield  {journal} {\bibinfo  {journal} {Phys.~Rev.~B}\ }\textbf {\bibinfo
  {volume} {97}},\ \bibinfo {pages} {121201} (\bibinfo {year}
  {2018})}\BibitemShut {NoStop}%
\bibitem [{\citenamefont {Lu}\ \emph {et~al.}(2019)\citenamefont {Lu},
  \citenamefont {Zhou},\ and\ \citenamefont {Bernardi}}]{Lu2019}%
  \BibitemOpen
  \bibfield  {author} {\bibinfo {author} {\bibfnamefont {I.-T.}\ \bibnamefont
  {Lu}}, \bibinfo {author} {\bibfnamefont {J.-J.}\ \bibnamefont {Zhou}}, \ and\
  \bibinfo {author} {\bibfnamefont {M.}~\bibnamefont {Bernardi}},\ }\href
  {\doibase 10.1103/PhysRevMaterials.3.033804} {\bibfield  {journal} {\bibinfo
  {journal} {Phys.~Rev.~Mat.}\ }\textbf {\bibinfo {volume} {3}},\ \bibinfo
  {pages} {033804} (\bibinfo {year} {2019})}\BibitemShut {NoStop}%
\bibitem [{\citenamefont {Lu}\ \emph {et~al.}(2020)\citenamefont {Lu},
  \citenamefont {Park}, \citenamefont {Zhou},\ and\ \citenamefont
  {Bernardi}}]{Lu2020}%
  \BibitemOpen
  \bibfield  {author} {\bibinfo {author} {\bibfnamefont {I.-T.}\ \bibnamefont
  {Lu}}, \bibinfo {author} {\bibfnamefont {J.}~\bibnamefont {Park}}, \bibinfo
  {author} {\bibfnamefont {J.-J.}\ \bibnamefont {Zhou}}, \ and\ \bibinfo
  {author} {\bibfnamefont {M.}~\bibnamefont {Bernardi}},\ }\href {\doibase
  https://doi.org/10.1038/s41524-020-0284-y} {\bibfield  {journal} {\bibinfo
  {journal} {npj~Comput.~Mater.}\ }\textbf {\bibinfo {volume} {6}},\ \bibinfo
  {pages} {1} (\bibinfo {year} {2020})}\BibitemShut {NoStop}%
\bibitem [{\citenamefont {Lu}\ \emph {et~al.}(2022)\citenamefont {Lu},
  \citenamefont {Zhou}, \citenamefont {Park},\ and\ \citenamefont
  {Bernardi}}]{lu2022first}%
  \BibitemOpen
  \bibfield  {author} {\bibinfo {author} {\bibfnamefont {I.-T.}\ \bibnamefont
  {Lu}}, \bibinfo {author} {\bibfnamefont {J.-J.}\ \bibnamefont {Zhou}},
  \bibinfo {author} {\bibfnamefont {J.}~\bibnamefont {Park}}, \ and\ \bibinfo
  {author} {\bibfnamefont {M.}~\bibnamefont {Bernardi}},\ }\href {\doibase
  https://doi.org/10.1103/PhysRevMaterials.6.L010801} {\bibfield  {journal}
  {\bibinfo  {journal} {Phys.~Rev.~Mat.}\ }\textbf {\bibinfo {volume} {6}},\
  \bibinfo {pages} {L010801} (\bibinfo {year} {2022})}\BibitemShut {NoStop}%
\bibitem [{\citenamefont {Ziman}(1972)}]{Ziman1972}%
  \BibitemOpen
  \bibfield  {author} {\bibinfo {author} {\bibfnamefont {J.~M.}\ \bibnamefont
  {Ziman}},\ }\href {\doibase 10.1017/CBO9781139644075} {\emph {\bibinfo
  {title} {{Principles of the theory of solids}}}},\ \bibinfo {edition} {2nd}\
  ed.\ (\bibinfo  {publisher} {Cambridge University Press},\ \bibinfo {year}
  {1972})\BibitemShut {NoStop}%
\bibitem [{\citenamefont {Blatt}(1957)}]{Blatt1957}%
  \BibitemOpen
  \bibfield  {author} {\bibinfo {author} {\bibfnamefont {F.~J.}\ \bibnamefont
  {Blatt}},\ }in\ \href {\doibase
  https://doi.org/10.1016/S0081-1947(08)60155-1} {\emph {\bibinfo {booktitle}
  {Solid~State~Phys.}}},\ Vol.~\bibinfo {volume} {4}\ (\bibinfo  {publisher}
  {Elsevier},\ \bibinfo {year} {1957})\ pp.\ \bibinfo {pages}
  {199--366}\BibitemShut {NoStop}%
\bibitem [{\citenamefont {Giustino}(2017)}]{Giustino2017}%
  \BibitemOpen
  \bibfield  {author} {\bibinfo {author} {\bibfnamefont {F.}~\bibnamefont
  {Giustino}},\ }\href {\doibase 10.1103/RevModPhys.89.015003} {\bibfield
  {journal} {\bibinfo  {journal} {Rev.~Mod.~Phys.}\ }\textbf {\bibinfo {volume}
  {89}},\ \bibinfo {pages} {015003} (\bibinfo {year} {2017})}\BibitemShut
  {NoStop}%
\bibitem [{\citenamefont {Ponc{\'e}}\ \emph {et~al.}(2021)\citenamefont
  {Ponc{\'e}}, \citenamefont {Macheda}, \citenamefont {Margine}, \citenamefont
  {Marzari}, \citenamefont {Bonini},\ and\ \citenamefont
  {Giustino}}]{ponce2021first}%
  \BibitemOpen
  \bibfield  {author} {\bibinfo {author} {\bibfnamefont {S.}~\bibnamefont
  {Ponc{\'e}}}, \bibinfo {author} {\bibfnamefont {F.}~\bibnamefont {Macheda}},
  \bibinfo {author} {\bibfnamefont {E.~R.}\ \bibnamefont {Margine}}, \bibinfo
  {author} {\bibfnamefont {N.}~\bibnamefont {Marzari}}, \bibinfo {author}
  {\bibfnamefont {N.}~\bibnamefont {Bonini}}, \ and\ \bibinfo {author}
  {\bibfnamefont {F.}~\bibnamefont {Giustino}},\ }\href {\doibase
  https://doi.org/10.1103/PhysRevResearch.3.043022} {\bibfield  {journal}
  {\bibinfo  {journal} {Phys.~Rev.~Research}\ }\textbf {\bibinfo {volume}
  {3}},\ \bibinfo {pages} {043022} (\bibinfo {year} {2021})}\BibitemShut
  {NoStop}%
\bibitem [{\citenamefont {Ma}\ \emph {et~al.}(2018)\citenamefont {Ma},
  \citenamefont {Nissimagoudar},\ and\ \citenamefont {Li}}]{Ma2018}%
  \BibitemOpen
  \bibfield  {author} {\bibinfo {author} {\bibfnamefont {J.}~\bibnamefont
  {Ma}}, \bibinfo {author} {\bibfnamefont {A.~S.}\ \bibnamefont
  {Nissimagoudar}}, \ and\ \bibinfo {author} {\bibfnamefont {W.}~\bibnamefont
  {Li}},\ }\href {\doibase 10.1103/PhysRevB.97.045201} {\bibfield  {journal}
  {\bibinfo  {journal} {Phys.~Rev.~B}\ }\textbf {\bibinfo {volume} {97}},\
  \bibinfo {pages} {045201} (\bibinfo {year} {2018})}\BibitemShut {NoStop}%
\bibitem [{\citenamefont {Li}(2015)}]{Li2015}%
  \BibitemOpen
  \bibfield  {author} {\bibinfo {author} {\bibfnamefont {W.}~\bibnamefont
  {Li}},\ }\href {\doibase 10.1103/PhysRevB.92.075405} {\bibfield  {journal}
  {\bibinfo  {journal} {Phys.~Rev.~B}\ }\textbf {\bibinfo {volume} {92}},\
  \bibinfo {pages} {075405} (\bibinfo {year} {2015})}\BibitemShut {NoStop}%
\bibitem [{\citenamefont {Madsen}\ and\ \citenamefont
  {Singh}(2006)}]{Madsen2006}%
  \BibitemOpen
  \bibfield  {author} {\bibinfo {author} {\bibfnamefont {G.~K.~H.}\
  \bibnamefont {Madsen}}\ and\ \bibinfo {author} {\bibfnamefont {D.~J.}\
  \bibnamefont {Singh}},\ }\href {\doibase 10.1016/j.cpc.2006.03.007}
  {\bibfield  {journal} {\bibinfo  {journal} {Comput.~Phys.~Commun.}\ }\textbf
  {\bibinfo {volume} {175}},\ \bibinfo {pages} {67} (\bibinfo {year}
  {2006})}\BibitemShut {NoStop}%
\bibitem [{\citenamefont {Madsen}\ \emph {et~al.}(2018)\citenamefont {Madsen},
  \citenamefont {Carrete},\ and\ \citenamefont {Verstraete}}]{Madsen2018}%
  \BibitemOpen
  \bibfield  {author} {\bibinfo {author} {\bibfnamefont {G.~K.~H.}\
  \bibnamefont {Madsen}}, \bibinfo {author} {\bibfnamefont {J.}~\bibnamefont
  {Carrete}}, \ and\ \bibinfo {author} {\bibfnamefont {M.~J.}\ \bibnamefont
  {Verstraete}},\ }\href {\doibase 10.1016/j.cpc.2018.05.010} {\bibfield
  {journal} {\bibinfo  {journal} {Comput.~Phys.~Commun.}\ }\textbf {\bibinfo
  {volume} {231}},\ \bibinfo {pages} {140} (\bibinfo {year}
  {2018})}\BibitemShut {NoStop}%
\bibitem [{\citenamefont {Hautier}\ \emph {et~al.}(2013)\citenamefont
  {Hautier}, \citenamefont {Miglio}, \citenamefont {Ceder}, \citenamefont
  {Rignanese},\ and\ \citenamefont {Gonze}}]{hautier2013identification}%
  \BibitemOpen
  \bibfield  {author} {\bibinfo {author} {\bibfnamefont {G.}~\bibnamefont
  {Hautier}}, \bibinfo {author} {\bibfnamefont {A.}~\bibnamefont {Miglio}},
  \bibinfo {author} {\bibfnamefont {G.}~\bibnamefont {Ceder}}, \bibinfo
  {author} {\bibfnamefont {G.-M.}\ \bibnamefont {Rignanese}}, \ and\ \bibinfo
  {author} {\bibfnamefont {X.}~\bibnamefont {Gonze}},\ }\href {\doibase
  https://doi.org/10.1038/ncomms3292} {\bibfield  {journal} {\bibinfo
  {journal} {Nat.~Commun.}\ }\textbf {\bibinfo {volume} {4}},\ \bibinfo {pages}
  {1} (\bibinfo {year} {2013})}\BibitemShut {NoStop}%
\bibitem [{\citenamefont {Ricci}\ \emph {et~al.}(2017)\citenamefont {Ricci},
  \citenamefont {Chen}, \citenamefont {Aydemir}, \citenamefont {Snyder},
  \citenamefont {Rignanese}, \citenamefont {Jain},\ and\ \citenamefont
  {Hautier}}]{ricci2017ab}%
  \BibitemOpen
  \bibfield  {author} {\bibinfo {author} {\bibfnamefont {F.}~\bibnamefont
  {Ricci}}, \bibinfo {author} {\bibfnamefont {W.}~\bibnamefont {Chen}},
  \bibinfo {author} {\bibfnamefont {U.}~\bibnamefont {Aydemir}}, \bibinfo
  {author} {\bibfnamefont {G.~J.}\ \bibnamefont {Snyder}}, \bibinfo {author}
  {\bibfnamefont {G.-M.}\ \bibnamefont {Rignanese}}, \bibinfo {author}
  {\bibfnamefont {A.}~\bibnamefont {Jain}}, \ and\ \bibinfo {author}
  {\bibfnamefont {G.}~\bibnamefont {Hautier}},\ }\href {\doibase
  https://doi.org/10.1038/sdata.2017.85} {\bibfield  {journal} {\bibinfo
  {journal} {Sci.~Data.}\ }\textbf {\bibinfo {volume} {4}},\ \bibinfo {pages}
  {1} (\bibinfo {year} {2017})}\BibitemShut {NoStop}%
\bibitem [{\citenamefont {Ricci}\ \emph {et~al.}(2020)\citenamefont {Ricci},
  \citenamefont {Dunn}, \citenamefont {Jain}, \citenamefont {Rignanese},\ and\
  \citenamefont {Hautier}}]{ricci2020gapped}%
  \BibitemOpen
  \bibfield  {author} {\bibinfo {author} {\bibfnamefont {F.}~\bibnamefont
  {Ricci}}, \bibinfo {author} {\bibfnamefont {A.}~\bibnamefont {Dunn}},
  \bibinfo {author} {\bibfnamefont {A.}~\bibnamefont {Jain}}, \bibinfo {author}
  {\bibfnamefont {G.-M.}\ \bibnamefont {Rignanese}}, \ and\ \bibinfo {author}
  {\bibfnamefont {G.}~\bibnamefont {Hautier}},\ }\href {\doibase
  10.1039/d0ta05197g} {\bibfield  {journal} {\bibinfo  {journal}
  {J.~Mater.~Chem.~A}\ }\textbf {\bibinfo {volume} {8}},\ \bibinfo {pages}
  {17579} (\bibinfo {year} {2020})}\BibitemShut {NoStop}%
\bibitem [{\citenamefont {Faghaninia}\ \emph {et~al.}(2017)\citenamefont
  {Faghaninia}, \citenamefont {Yu}, \citenamefont {Aydemir}, \citenamefont
  {Wood}, \citenamefont {Chen}, \citenamefont {Rignanese}, \citenamefont
  {Snyder}, \citenamefont {Hautier},\ and\ \citenamefont
  {Jain}}]{faghaninia2017computational}%
  \BibitemOpen
  \bibfield  {author} {\bibinfo {author} {\bibfnamefont {A.}~\bibnamefont
  {Faghaninia}}, \bibinfo {author} {\bibfnamefont {G.}~\bibnamefont {Yu}},
  \bibinfo {author} {\bibfnamefont {U.}~\bibnamefont {Aydemir}}, \bibinfo
  {author} {\bibfnamefont {M.}~\bibnamefont {Wood}}, \bibinfo {author}
  {\bibfnamefont {W.}~\bibnamefont {Chen}}, \bibinfo {author} {\bibfnamefont
  {G.-M.}\ \bibnamefont {Rignanese}}, \bibinfo {author} {\bibfnamefont {G.~J.}\
  \bibnamefont {Snyder}}, \bibinfo {author} {\bibfnamefont {G.}~\bibnamefont
  {Hautier}}, \ and\ \bibinfo {author} {\bibfnamefont {A.}~\bibnamefont
  {Jain}},\ }\href {\doibase https://doi.org/10.1039/C7CP00437K} {\bibfield
  {journal} {\bibinfo  {journal} {Phys.~Chem.~Chem.~Phys.}\ }\textbf {\bibinfo
  {volume} {19}},\ \bibinfo {pages} {6743} (\bibinfo {year}
  {2017})}\BibitemShut {NoStop}%
\bibitem [{\citenamefont {Brunin}\ \emph {et~al.}(2019)\citenamefont {Brunin},
  \citenamefont {Ricci}, \citenamefont {Ha}, \citenamefont {Rignanese},\ and\
  \citenamefont {Hautier}}]{brunin2019transparent}%
  \BibitemOpen
  \bibfield  {author} {\bibinfo {author} {\bibfnamefont {G.}~\bibnamefont
  {Brunin}}, \bibinfo {author} {\bibfnamefont {F.}~\bibnamefont {Ricci}},
  \bibinfo {author} {\bibfnamefont {V.-A.}\ \bibnamefont {Ha}}, \bibinfo
  {author} {\bibfnamefont {G.-M.}\ \bibnamefont {Rignanese}}, \ and\ \bibinfo
  {author} {\bibfnamefont {G.}~\bibnamefont {Hautier}},\ }\href {\doibase
  https://doi.org/10.1038/s41524-019-0200-5} {\bibfield  {journal} {\bibinfo
  {journal} {npj~Comput.~Mater.}\ }\textbf {\bibinfo {volume} {5}},\ \bibinfo
  {pages} {1} (\bibinfo {year} {2019})}\BibitemShut {NoStop}%
\bibitem [{\citenamefont {Gibbs}\ \emph {et~al.}(2017)\citenamefont {Gibbs},
  \citenamefont {Ricci}, \citenamefont {Li}, \citenamefont {Zhu}, \citenamefont
  {Persson}, \citenamefont {Ceder}, \citenamefont {Hautier}, \citenamefont
  {Jain},\ and\ \citenamefont {Snyder}}]{gibbs2017effective}%
  \BibitemOpen
  \bibfield  {author} {\bibinfo {author} {\bibfnamefont {Z.~M.}\ \bibnamefont
  {Gibbs}}, \bibinfo {author} {\bibfnamefont {F.}~\bibnamefont {Ricci}},
  \bibinfo {author} {\bibfnamefont {G.}~\bibnamefont {Li}}, \bibinfo {author}
  {\bibfnamefont {H.}~\bibnamefont {Zhu}}, \bibinfo {author} {\bibfnamefont
  {K.}~\bibnamefont {Persson}}, \bibinfo {author} {\bibfnamefont
  {G.}~\bibnamefont {Ceder}}, \bibinfo {author} {\bibfnamefont
  {G.}~\bibnamefont {Hautier}}, \bibinfo {author} {\bibfnamefont
  {A.}~\bibnamefont {Jain}}, \ and\ \bibinfo {author} {\bibfnamefont {G.~J.}\
  \bibnamefont {Snyder}},\ }\href {\doibase
  https://doi.org/10.1038/s41524-017-0013-3} {\bibfield  {journal} {\bibinfo
  {journal} {npj~Comput.~Mater.}\ }\textbf {\bibinfo {volume} {3}},\ \bibinfo
  {pages} {1} (\bibinfo {year} {2017})}\BibitemShut {NoStop}%
\bibitem [{\citenamefont {Smiadak}\ \emph {et~al.}(2022)\citenamefont
  {Smiadak}, \citenamefont {Claes}, \citenamefont {Perez}, \citenamefont
  {Marshall}, \citenamefont {Peng}, \citenamefont {Chen}, \citenamefont
  {Hautier}, \citenamefont {Schierning},\ and\ \citenamefont
  {Zevalkink}}]{smiadak2022quasi}%
  \BibitemOpen
  \bibfield  {author} {\bibinfo {author} {\bibfnamefont {D.~M.}\ \bibnamefont
  {Smiadak}}, \bibinfo {author} {\bibfnamefont {R.}~\bibnamefont {Claes}},
  \bibinfo {author} {\bibfnamefont {N.}~\bibnamefont {Perez}}, \bibinfo
  {author} {\bibfnamefont {M.}~\bibnamefont {Marshall}}, \bibinfo {author}
  {\bibfnamefont {W.}~\bibnamefont {Peng}}, \bibinfo {author} {\bibfnamefont
  {W.}~\bibnamefont {Chen}}, \bibinfo {author} {\bibfnamefont {G.}~\bibnamefont
  {Hautier}}, \bibinfo {author} {\bibfnamefont {G.}~\bibnamefont {Schierning}},
  \ and\ \bibinfo {author} {\bibfnamefont {A.}~\bibnamefont {Zevalkink}},\
  }\href {\doibase https://doi.org/10.1016/j.mtphys.2021.100597} {\bibfield
  {journal} {\bibinfo  {journal} {Mater.~Today~Phys.}\ ,\ \bibinfo {pages}
  {100597}} (\bibinfo {year} {2022})}\BibitemShut {NoStop}%
\bibitem [{\citenamefont {Chmielowski}\ \emph {et~al.}(2015)\citenamefont
  {Chmielowski}, \citenamefont {P{\'e}r{\'e}}, \citenamefont {Bera},
  \citenamefont {Opahle}, \citenamefont {Xie}, \citenamefont {Jacob},
  \citenamefont {Capet}, \citenamefont {Roussel}, \citenamefont {Weidenkaff},
  \citenamefont {Madsen} \emph {et~al.}}]{chmielowski2015theoretical}%
  \BibitemOpen
  \bibfield  {author} {\bibinfo {author} {\bibfnamefont {R.}~\bibnamefont
  {Chmielowski}}, \bibinfo {author} {\bibfnamefont {D.}~\bibnamefont
  {P{\'e}r{\'e}}}, \bibinfo {author} {\bibfnamefont {C.}~\bibnamefont {Bera}},
  \bibinfo {author} {\bibfnamefont {I.}~\bibnamefont {Opahle}}, \bibinfo
  {author} {\bibfnamefont {W.}~\bibnamefont {Xie}}, \bibinfo {author}
  {\bibfnamefont {S.}~\bibnamefont {Jacob}}, \bibinfo {author} {\bibfnamefont
  {F.}~\bibnamefont {Capet}}, \bibinfo {author} {\bibfnamefont
  {P.}~\bibnamefont {Roussel}}, \bibinfo {author} {\bibfnamefont
  {A.}~\bibnamefont {Weidenkaff}}, \bibinfo {author} {\bibfnamefont {G.~K.}\
  \bibnamefont {Madsen}},  \emph {et~al.},\ }\href {\doibase
  doi:10.1063/1.4916528} {\bibfield  {journal} {\bibinfo  {journal}
  {J.~Appl.~Phys.}\ }\textbf {\bibinfo {volume} {117}},\ \bibinfo {pages}
  {125103} (\bibinfo {year} {2015})}\BibitemShut {NoStop}%
\bibitem [{\citenamefont {Petretto}\ \emph
  {et~al.}(2018{\natexlab{a}})\citenamefont {Petretto}, \citenamefont
  {Dwaraknath}, \citenamefont {Miranda}, \citenamefont {Winston}, \citenamefont
  {Giantomassi}, \citenamefont {{Van Setten}}, \citenamefont {Gonze},
  \citenamefont {Persson}, \citenamefont {Hautier},\ and\ \citenamefont
  {Rignanese}}]{Petretto2018}%
  \BibitemOpen
  \bibfield  {author} {\bibinfo {author} {\bibfnamefont {G.}~\bibnamefont
  {Petretto}}, \bibinfo {author} {\bibfnamefont {S.}~\bibnamefont
  {Dwaraknath}}, \bibinfo {author} {\bibfnamefont {H.~P.~C.}\ \bibnamefont
  {Miranda}}, \bibinfo {author} {\bibfnamefont {D.}~\bibnamefont {Winston}},
  \bibinfo {author} {\bibfnamefont {M.}~\bibnamefont {Giantomassi}}, \bibinfo
  {author} {\bibfnamefont {M.~J.}\ \bibnamefont {{Van Setten}}}, \bibinfo
  {author} {\bibfnamefont {X.}~\bibnamefont {Gonze}}, \bibinfo {author}
  {\bibfnamefont {K.~A.}\ \bibnamefont {Persson}}, \bibinfo {author}
  {\bibfnamefont {G.}~\bibnamefont {Hautier}}, \ and\ \bibinfo {author}
  {\bibfnamefont {G.~M.}\ \bibnamefont {Rignanese}},\ }\href {\doibase
  10.1038/sdata.2018.65} {\bibfield  {journal} {\bibinfo  {journal}
  {Sci.~Data.}\ }\textbf {\bibinfo {volume} {5}},\ \bibinfo {pages} {1}
  (\bibinfo {year} {2018}{\natexlab{a}})}\BibitemShut {NoStop}%
\bibitem [{\citenamefont {Jain}\ \emph {et~al.}(2013)\citenamefont {Jain},
  \citenamefont {Ong}, \citenamefont {Hautier}, \citenamefont {Chen},
  \citenamefont {Richards}, \citenamefont {Dacek}, \citenamefont {Cholia},
  \citenamefont {Gunter}, \citenamefont {Skinner}, \citenamefont {Ceder} \emph
  {et~al.}}]{jain2013commentary}%
  \BibitemOpen
  \bibfield  {author} {\bibinfo {author} {\bibfnamefont {A.}~\bibnamefont
  {Jain}}, \bibinfo {author} {\bibfnamefont {S.~P.}\ \bibnamefont {Ong}},
  \bibinfo {author} {\bibfnamefont {G.}~\bibnamefont {Hautier}}, \bibinfo
  {author} {\bibfnamefont {W.}~\bibnamefont {Chen}}, \bibinfo {author}
  {\bibfnamefont {W.~D.}\ \bibnamefont {Richards}}, \bibinfo {author}
  {\bibfnamefont {S.}~\bibnamefont {Dacek}}, \bibinfo {author} {\bibfnamefont
  {S.}~\bibnamefont {Cholia}}, \bibinfo {author} {\bibfnamefont
  {D.}~\bibnamefont {Gunter}}, \bibinfo {author} {\bibfnamefont
  {D.}~\bibnamefont {Skinner}}, \bibinfo {author} {\bibfnamefont
  {G.}~\bibnamefont {Ceder}},  \emph {et~al.},\ }\href {\doibase
  https://doi.org/10.1063/1.4812323} {\bibfield  {journal} {\bibinfo  {journal}
  {APL~Mater.}\ }\textbf {\bibinfo {volume} {1}},\ \bibinfo {pages} {011002}
  (\bibinfo {year} {2013})}\BibitemShut {NoStop}%
\bibitem [{\citenamefont {van Setten}\ \emph {et~al.}(2018)\citenamefont {van
  Setten}, \citenamefont {Giantomassi}, \citenamefont {Bousquet}, \citenamefont
  {Verstraete}, \citenamefont {Hamann}, \citenamefont {Gonze},\ and\
  \citenamefont {Rignanese}}]{van2018pseudodojo}%
  \BibitemOpen
  \bibfield  {author} {\bibinfo {author} {\bibfnamefont {M.~J.}\ \bibnamefont
  {van Setten}}, \bibinfo {author} {\bibfnamefont {M.}~\bibnamefont
  {Giantomassi}}, \bibinfo {author} {\bibfnamefont {E.}~\bibnamefont
  {Bousquet}}, \bibinfo {author} {\bibfnamefont {M.~J.}\ \bibnamefont
  {Verstraete}}, \bibinfo {author} {\bibfnamefont {D.~R.}\ \bibnamefont
  {Hamann}}, \bibinfo {author} {\bibfnamefont {X.}~\bibnamefont {Gonze}}, \
  and\ \bibinfo {author} {\bibfnamefont {G.-M.}\ \bibnamefont {Rignanese}},\
  }\href {\doibase https://doi.org/10.1016/j.cpc.2018.01.012} {\bibfield
  {journal} {\bibinfo  {journal} {Comput.~Phys.~Commun.}\ }\textbf {\bibinfo
  {volume} {226}},\ \bibinfo {pages} {39} (\bibinfo {year} {2018})}\BibitemShut
  {NoStop}%
\bibitem [{Note1()}]{Note1}%
  \BibitemOpen
  \bibinfo {note} {Dynamical quadrupoles are non-zero in all
  non-centrosymmetric crystals, but also in centrosymmetric ones if one or more
  atoms are placed at non-centrosymmetric sites.}\BibitemShut {Stop}%
\bibitem [{\citenamefont {Jhalani}\ \emph {et~al.}(2020)\citenamefont
  {Jhalani}, \citenamefont {Zhou}, \citenamefont {Park}, \citenamefont
  {Dreyer},\ and\ \citenamefont {Bernardi}}]{jhalani2020piezoelectric}%
  \BibitemOpen
  \bibfield  {author} {\bibinfo {author} {\bibfnamefont {V.~A.}\ \bibnamefont
  {Jhalani}}, \bibinfo {author} {\bibfnamefont {J.-J.}\ \bibnamefont {Zhou}},
  \bibinfo {author} {\bibfnamefont {J.}~\bibnamefont {Park}}, \bibinfo {author}
  {\bibfnamefont {C.~E.}\ \bibnamefont {Dreyer}}, \ and\ \bibinfo {author}
  {\bibfnamefont {M.}~\bibnamefont {Bernardi}},\ }\href {\doibase
  10.1103/PhysRevLett.125.136602} {\bibfield  {journal} {\bibinfo  {journal}
  {Phys.~Rev.~Lett.}\ }\textbf {\bibinfo {volume} {125}},\ \bibinfo {pages}
  {136602} (\bibinfo {year} {2020})}\BibitemShut {NoStop}%
\bibitem [{\citenamefont {Park}\ \emph {et~al.}(2020)\citenamefont {Park},
  \citenamefont {Zhou}, \citenamefont {Jhalani}, \citenamefont {Dreyer},\ and\
  \citenamefont {Bernardi}}]{park2020long}%
  \BibitemOpen
  \bibfield  {author} {\bibinfo {author} {\bibfnamefont {J.}~\bibnamefont
  {Park}}, \bibinfo {author} {\bibfnamefont {J.-J.}\ \bibnamefont {Zhou}},
  \bibinfo {author} {\bibfnamefont {V.~A.}\ \bibnamefont {Jhalani}}, \bibinfo
  {author} {\bibfnamefont {C.~E.}\ \bibnamefont {Dreyer}}, \ and\ \bibinfo
  {author} {\bibfnamefont {M.}~\bibnamefont {Bernardi}},\ }\href {\doibase
  https://doi.org/10.1103/PhysRevB.102.125203} {\bibfield  {journal} {\bibinfo
  {journal} {Phys.~Rev.~B}\ }\textbf {\bibinfo {volume} {102}},\ \bibinfo
  {pages} {125203} (\bibinfo {year} {2020})}\BibitemShut {NoStop}%
\bibitem [{\citenamefont {{See Supplemental Material at \url{http://link} for
  more information about computational details and results}}()}]{supplemental}%
  \BibitemOpen
  \bibfield  {author} {\bibinfo {author} {\bibnamefont {{See Supplemental
  Material at \url{http://link} for more information about computational
  details and results}}},\ }\href@noop {} {}\BibitemShut {NoStop}%
\bibitem [{\citenamefont {Zhou}\ and\ \citenamefont
  {Bernardi}(2016)}]{zhou2016ab}%
  \BibitemOpen
  \bibfield  {author} {\bibinfo {author} {\bibfnamefont {J.-J.}\ \bibnamefont
  {Zhou}}\ and\ \bibinfo {author} {\bibfnamefont {M.}~\bibnamefont
  {Bernardi}},\ }\href {\doibase https://doi.org/10.1103/PhysRevB.94.201201}
  {\bibfield  {journal} {\bibinfo  {journal} {Phys.~Rev.~B}\ }\textbf {\bibinfo
  {volume} {94}},\ \bibinfo {pages} {201201} (\bibinfo {year}
  {2016})}\BibitemShut {NoStop}%
\bibitem [{\citenamefont {Giantomassi~\textit{et al.}}()}]{abipy-website}%
  \BibitemOpen
  \bibfield  {author} {\bibinfo {author} {\bibfnamefont {M.}~\bibnamefont
  {Giantomassi~\textit{et al.}}},\ }\href@noop {} {\enquote {\bibinfo {title}
  {{AbiPy project}},}\ }\bibinfo {howpublished} {\url
  {https://github.com/abinit/abipy}}\BibitemShut {NoStop}%
\bibitem [{\citenamefont {Petretto}\ \emph
  {et~al.}(2018{\natexlab{b}})\citenamefont {Petretto}, \citenamefont {Gonze},
  \citenamefont {Hautier},\ and\ \citenamefont {Rignanese}}]{Petretto2018b}%
  \BibitemOpen
  \bibfield  {author} {\bibinfo {author} {\bibfnamefont {G.}~\bibnamefont
  {Petretto}}, \bibinfo {author} {\bibfnamefont {X.}~\bibnamefont {Gonze}},
  \bibinfo {author} {\bibfnamefont {G.}~\bibnamefont {Hautier}}, \ and\
  \bibinfo {author} {\bibfnamefont {G.-M.}\ \bibnamefont {Rignanese}},\ }\href
  {\doibase https://doi.org/10.1016/j.commatsci.2017.12.040} {\bibfield
  {journal} {\bibinfo  {journal} {Comput.~Mater.~Sci.}\ }\textbf {\bibinfo
  {volume} {144}},\ \bibinfo {pages} {331} (\bibinfo {year}
  {2018}{\natexlab{b}})}\BibitemShut {NoStop}%
\bibitem [{\citenamefont {Shankland}(1971)}]{Shankland1971}%
  \BibitemOpen
  \bibfield  {author} {\bibinfo {author} {\bibfnamefont {D.~G.}\ \bibnamefont
  {Shankland}},\ }\href {\doibase 10.1002/qua.560050857} {\bibfield  {journal}
  {\bibinfo  {journal} {Int.~J.~Quantum Chem.}\ }\textbf {\bibinfo {volume}
  {5}},\ \bibinfo {pages} {497} (\bibinfo {year} {1971})}\BibitemShut {NoStop}%
\bibitem [{\citenamefont {Euwema}\ \emph {et~al.}(1969)\citenamefont {Euwema},
  \citenamefont {Stukel}, \citenamefont {Collins}, \citenamefont {DeWitt},\
  and\ \citenamefont {Shankland}}]{Euwema1969}%
  \BibitemOpen
  \bibfield  {author} {\bibinfo {author} {\bibfnamefont {R.~N.}\ \bibnamefont
  {Euwema}}, \bibinfo {author} {\bibfnamefont {D.~J.}\ \bibnamefont {Stukel}},
  \bibinfo {author} {\bibfnamefont {T.~C.}\ \bibnamefont {Collins}}, \bibinfo
  {author} {\bibfnamefont {J.~S.}\ \bibnamefont {DeWitt}}, \ and\ \bibinfo
  {author} {\bibfnamefont {D.~G.}\ \bibnamefont {Shankland}},\ }\href {\doibase
  10.1103/PhysRev.178.1419} {\bibfield  {journal} {\bibinfo  {journal}
  {Phys.~Rev.}\ }\textbf {\bibinfo {volume} {178}},\ \bibinfo {pages} {1419}
  (\bibinfo {year} {1969})}\BibitemShut {NoStop}%
\bibitem [{\citenamefont {Koelling}\ and\ \citenamefont
  {Wood}(1986)}]{Koelling1986}%
  \BibitemOpen
  \bibfield  {author} {\bibinfo {author} {\bibfnamefont {D.~D.}\ \bibnamefont
  {Koelling}}\ and\ \bibinfo {author} {\bibfnamefont {J.~H.}\ \bibnamefont
  {Wood}},\ }\href {\doibase 10.1016/0021-9991(86)90261-5} {\bibfield
  {journal} {\bibinfo  {journal} {J.~Comput.~Phys.}\ }\textbf {\bibinfo
  {volume} {67}},\ \bibinfo {pages} {253} (\bibinfo {year} {1986})}\BibitemShut
  {NoStop}%
\end{thebibliography}%

\end{document}